\documentclass{DISproc}

\usepackage{amsmath}
\usepackage{wrapfig}

\newcommand{\bi} {\begin{itemize}}
\newcommand{\ei} {\end{itemize}}
\newcommand{\beq}{\begin{equation}}
\newcommand{\eeq}{\end{equation}}
\newcommand{\bea}{\begin{eqnarray}}
\newcommand{\eea}{\end{eqnarray}}

\begin{document}
\title{The NLO jet vertex for Mueller-Navelet and forward jets in the 
small-cone approximation}

\author{{\slshape Dmitry Yu. Ivanov$^1$, Alessandro Papa$^2$}\\[1ex]
$^1$Sobolev Institute of Mathematics and Novosibirsk State University,
630090 Novosibirsk, Russia\\
$^2$Dipartimento di Fisica, Universit\`a della Calabria,
and Istituto Nazionale di Fisica Nucleare, Gruppo collegato di Cosenza,
Arcavacata di Rende, I-87036 Cosenza, Italy}

\contribID{xy}

\doi  

\maketitle

\begin{abstract}
We calculate in the next-to-leading order the impact factor (vertex) for 
the production of a forward high-$p_T$ jet, in the approximation of small 
aperture of the jet cone in the pseudorapidity-azimuthal angle plane. The 
final expression for the vertex turns out to be simple and easy to implement 
in numerical calculations.
\end{abstract}

\section{Introduction}

\begin{wrapfigure}{R}{0.25\textwidth}
\centering
\includegraphics[width=0.25\textwidth]{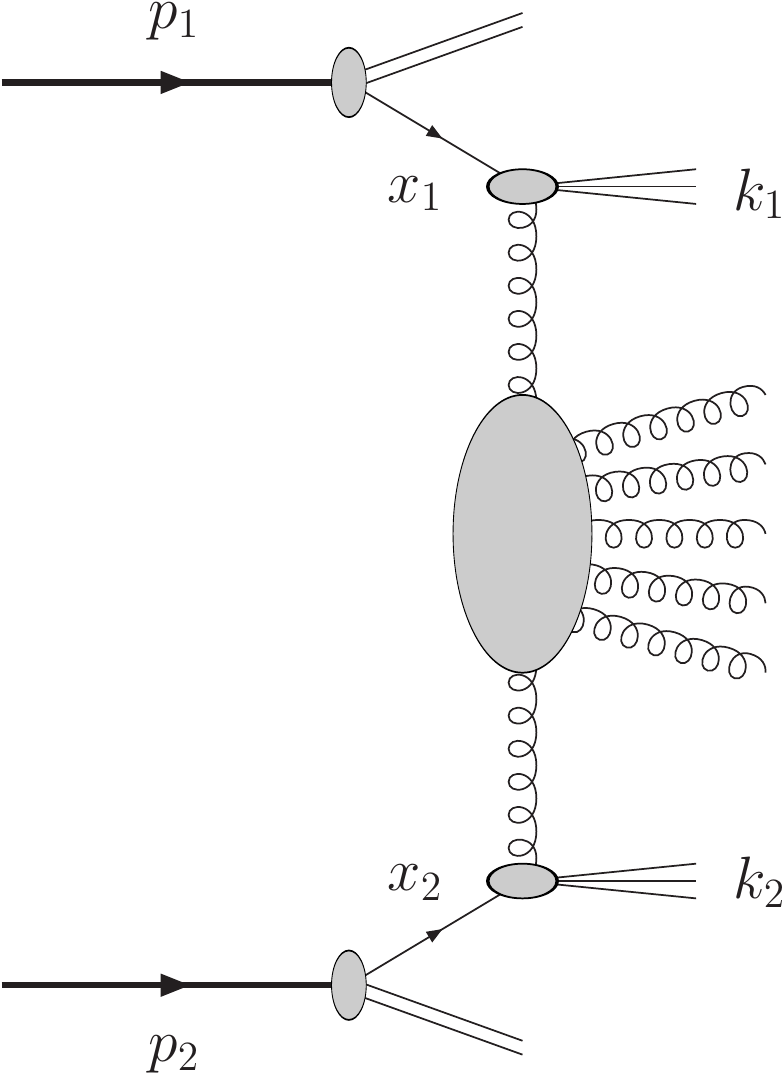}
\caption{Schematic representation of the Mueller-Navelet jet production.}
\label{fig1}
\end{wrapfigure}

We consider the process $\rm{proton}(p_1) + \rm{proton}(p_2) \to  
\rm{jet}(k_1) + \rm{jet}(k_2) + X$. Introducing the Sudakov decomposition
($s=2 p_1\cdot p_2$),
\[
k_1= \alpha_1 p_1+ \frac{\vec k_1^{\:2}}{\alpha_1 s}p_2+k_{1,\perp} \;, 
\;\;\; k_{1,\perp}^2=-\vec k_1^{\:2} \;,
\]
\[
k_2= \alpha_2 p_2+ \frac{\vec k_2^{\:2}}{\alpha_2 s}p_1+k_{2,\perp} \;, 
\;\;\; k_{2,\perp}^2=-\vec k_2^{\:2}\;,
\]
we take the kinematics when jet transverse momenta are large,
$\vec k_1^{\:2}\sim \vec k_2^{\:2} \gg \Lambda_{\rm QCD}^2$, and 
there is a large rapidity gap between jets, 
$\Delta y = \ln \frac{\alpha_1 \alpha_2 s}{|\vec k_1| |\vec k_2|}$,
which requires large c.m. energy of the proton collisions,
$s=2 p_1\cdot p_2 \gg \vec k_{1,2}^{\:2}$. In the perturbative QCD 
description of the process, the hard scale is provided by the 
jet transverse momenta, $Q^2 \sim \vec k_{1,2}^2 \gg \Lambda_{\rm QCD}^2$; 
moreover, we neglect power-suppressed contributions $\sim 1/Q$, thus 
allowing the use of leading-twist PDFs, $f_g(x)$ and $f_q(x)$. We still
need to resum the QCD perturbative series, according to DGLAP~\cite{DGLAP},
$\sum_n a_n (\alpha_s^n\ln^n Q^2 + b_n \alpha_s^n\ln^{n-1} Q^2)$ and 
BFKL~\cite{BFKL}, $\sum_n (c_n \alpha_s^n\ln^n s + d_n \alpha_s^n\ln^{n-1} s)$.
Mueller and Navelet~\cite{Mueller:1986ey} proposed that, for $\Delta y \gg 1$, 
the BFKL approach is more adequate and leads to a \emph{faster} energy 
dependence and more decorrelation in the relative jet azimuthal angle
$\phi=\phi_1-\phi_2-\pi$.

In the BFKL approach, valid in the Regge limit $s \rightarrow 
\infty$, the total cross section of a hard process $A + B \to X$, via the 
optical theorem, $\sigma = \frac{\rm{Im}_s {\cal A}}{s}$, can be written 
as the convolution of the Green's function of two interacting 
Reggeized gluons and of the impact factors of the colliding particles.
This is valid both in the LLA (resummation of all terms $(\alpha_s\ln s)^n$)
and in the NLA (resummation of all terms $\alpha_s(\alpha_s\ln s)^n$).
In formulae, 
\[
{\rm Im}_s {\cal A}=\frac{s}{(2\pi)^{D-2}}\!\int\!\frac{d^{D-2}\vec q_1}{\vec
q_1^{\,\, 2}}\Phi_A(\vec q_1,s_0)\!\int\!
\frac{d^{D-2}\vec q_2}{\vec q_2^{\,\,2}} \Phi_B(-\vec q_2,s_0)
\!\int\limits^{\delta +i\infty}_{\delta
-i\infty}\!\frac{d\omega}{2\pi i}\left(\frac{s}{s_0}
\right)^\omega G_\omega (\vec q_1, \vec q_2)\;.
\]
The Green's function is process-independent and is determined through the 
BFKL equation,
\[
\omega \, G_\omega (\vec q_1,\vec q_2)  =\delta^{D-2} (\vec q_1-\vec q_2)
+\int d^{D-2}\vec q \, K(\vec q_1,\vec q) \,G_\omega (\vec q, \vec q_1)\;,
\]
whereas impact factors are process-dependent and only very few of them
have been calculated in the NLA. For the process under consideration, the 
starting point is provided by the impact factors for colliding 
partons~\cite{FFKP99,Ciafaloni:2000sq} (see Fig.~2).

\begin{figure}[htb]
\centering
\includegraphics[width=0.35\textwidth]{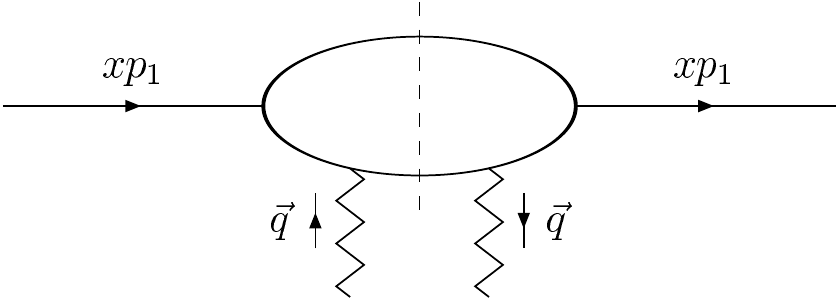}
\hspace{1cm}
\includegraphics[width=0.35\textwidth]{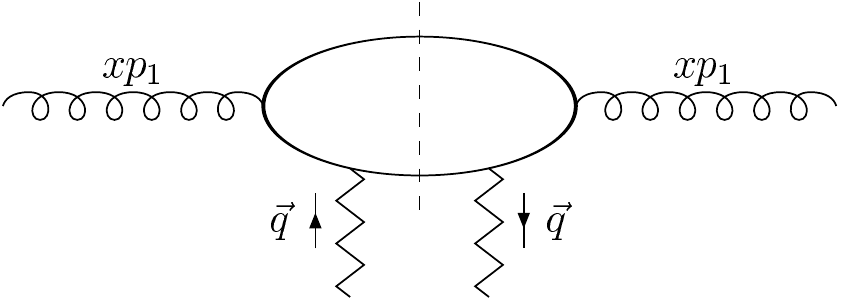}
\caption{Schematic representation of quark (left) and gluon (right)
impact factors.}
\label{fig2}
\end{figure}

In the LLA one needs leading-order (LO) impact factors, which take contribution
only from a one-particle intermediate state in the parton-Reggeon collision; in
the NLA one needs next-to-LO (NLO) impact factors, which take contributions
from virtual corrections (one-particle intermediate states) and real 
particle production (two-particle intermediate states). The steps to get
the quark(gluon) jet vertex from the quark(gluon) parton impact factor
are: (1) \emph{open} one of the integrations over the phase space of the 
intermediate state to \emph{allow} one parton to generate the jet, (2)
take the convolution with PDFs, $\sum_{a=q,\bar q} f_a \otimes 
({\rm quark \ jet \ vertex}) + f_g \otimes ({\rm gluon \ jet \ vertex})$,
(3) project onto the eigenfunctions of the LO BFKL kernel, i.e. transfer to 
the $(\nu,n)$-representation
\[
\Phi(\nu,n)=\int d^2\vec q \,\frac{\Phi(\vec q)}{\vec q^{\,\, 2}}\frac{1}{\pi
\sqrt{2}}\left(\vec q^{\,\, 2}\right)^{\gamma-\frac{n}{2}} 
\left(\vec q \cdot \vec l \,\, \right)^n \;,
\;\;\;\;\;
\gamma=i\nu-\frac{1}{2}\;,\;\;\;\;\;\vec l^{\:2}=0\;.
\]
The NLO jet vertices have been calculated in the transverse momentum space 
(no step (3)) in~\cite{Bartels:2002yj} and cross-checked 
in~\cite{Caporale:2011cc}. They are given by complicated expression, to be 
transferred numerically to the $(\nu,n)$-representation, as it was done 
in~\cite{Colferai:2010wu}, where they were used to study Mueller-Navelet jets 
in the NLA with LHC kinematics.

Here we want to sketch the derivation of an approximated expression for
jet vertices, valid for jets with small aperture of the cone in the 
pseudorapidity - azimuthal angle plane. The details of the calculation
are given in~\cite{IP12}. 

\section{Jet definition, small-cone approximation (SCA) and outline of the calculation}

In the LO we have a one-particle intermediate state and the 
kinematics of the produced parton $a$ is completely fixed by the jet 
kinematics (see Fig.~3).

\begin{wrapfigure}{R}{0.2\textwidth}
\centering
\includegraphics[width=0.2\textwidth]{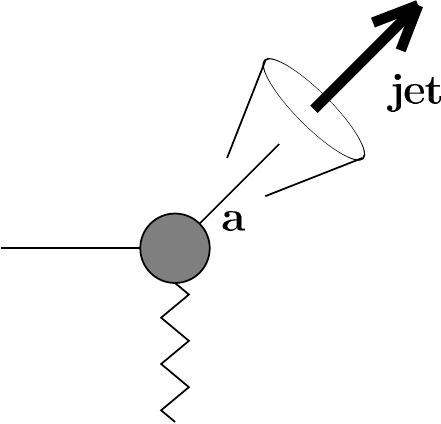}
\caption{Parton-Reggeon collision, the jet is formed by a single parton.}
\label{fig3}
\end{wrapfigure}

In the NLO, when real corrections are considered, we have two-particle
intermediate states. Then, we can have the following cases:
(i) the parton $a$ generates the jet, while the parton $b$ can have arbitrary
kinematics, provided that it lies \emph{outside} the jet cone; 
(ii) similarly with $a\leftrightarrow b$; (iii) the two partons $a$ and $b$ 
both generate the jet (see Fig.~4(left)).

The case in which one parton (say $a$) generates the jet and the other 
parton is outside the jet cone can also be written as the contribution 
when the parton $a$ is produced with the same jet kinematics while the 
parton $b$ can have any kinematics (\emph{inclusive} jet production by the
parton $a$) \emph{minus} the contribution when the parton $b$ lies 
\emph{inside} the jet cone (see Fig.~4(right)). 

The SCA~\cite{Jager:2004jh} means that all cones which appear in the jet
definition given above are to be taken with aperture (in the 
pseudorapidity - azimuthal angle plane) smaller than a fixed value $R$.
For $s\sim Q^2$, very good agreement between SCA and Monte Carlo calculations
was found for cone sizes up to $R=0.7$~\cite{Jager:2004jh}.

\begin{figure}[h]
\centering
\includegraphics[width=0.33\textwidth]{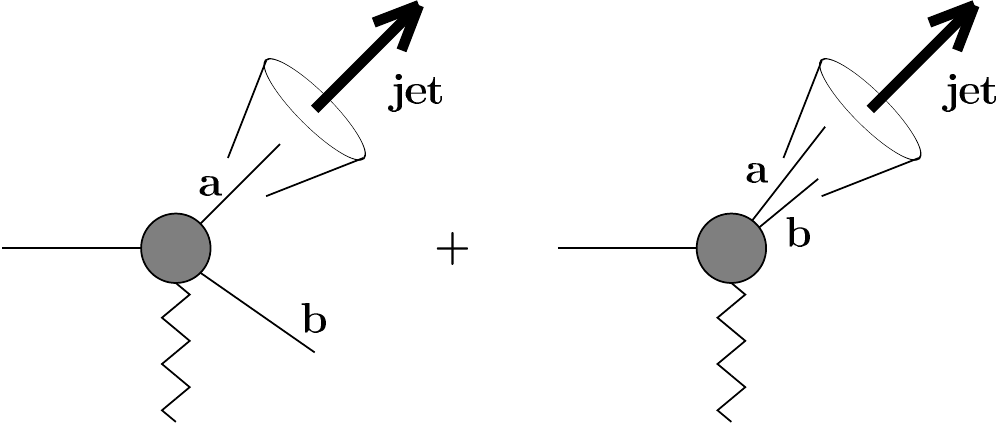}
\hspace{2cm}
\includegraphics[width=0.5\textwidth]{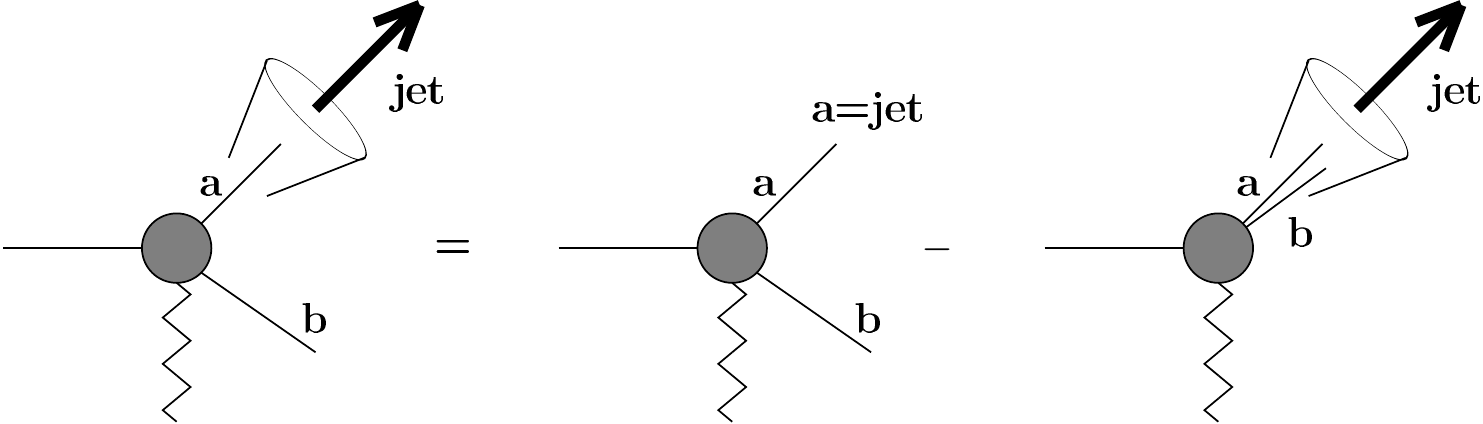}
\caption{(Left) Parton-Reggeon collision, two partons are produced and the 
jet is formed either by one of the partons or by both partons.
(Right) The production of the jet by one parton when the second one is
outside the cone can be seen as the \emph{inclusive} production minus the
contribution when the second parton is inside the cone.}
\label{fig4}
\end{figure}

For the jet vertex in the LLA the starting point is given by the 
\emph{inclusive} LO parton impact factors, $\Phi_q=g^2\frac{\sqrt{N^2-1}}{2N}$
and $\Phi_g=\frac{C_A}{C_F}\Phi_q$. Then we have to \emph{open} the 
integration over the one-particle intermediate state, i.e. introduce suitable 
delta functions, and take the convolution with the PDFs, getting
($\alpha$ and $\vec k$ are the jet kinematic variables)
\[
\frac{d\Phi^J}{\vec q^{\,\, 2}}={\cal C}\int \,d\alpha
\frac{d^2\vec k}{\vec k^{\,2}}\, dx \, \delta^{(2)}\left(\vec k-\vec q\right)
\delta(\alpha-x)\left(\frac{C_A}{C_F}f_g(x)+\sum_{a=q,\bar q} f_a(x)\right)\;.
\]
This expression can be used to construct the collinear and QCD coupling
counterterms in the NLA, arising when the renormalization of the PDFs and of 
the QCD coupling are taken into account.

For the jet vertex in the NLA, we separate the cases of quark- and 
gluon-iniziated subprocesses. For incoming quark, we have the following 
contributions: (a) virtual corrections, (b) real corrections from the 
quark-gluon state. The contribution (b) can be separated into 
the following pieces: (b$_1$) both quark and gluon generate the jet, 
(b$_2$) gluon \emph{inclusive} jet generation minus gluon \emph{inclusive}
jet generation with the quark in the jet cone, (b$_3$) quark \emph{inclusive}
jet generation minus quark \emph{inclusive} jet generation with the gluon 
in the jet cone. For incoming gluon, we have the following contributions:
(a) virtual corrections, (b) real corrections from quark-antiquark 
state, (c) real corrections from two-gluon state. The contribution (b) can be 
separated into the following pieces: (b$_1$) both quark and antiquark
generate the jet, (b$_2$) (anti)quark \emph{inclusive} jet generation
minus (anti)quark \emph{inclusive} jet generation with the antiquark(quark) 
in the jet cone. The contribution (c) can be separated into the following 
pieces: (c$_1$) both gluons generate the jet, (c$_2$) gluon \emph{inclusive} 
jet generation minus gluon \emph{inclusive} jet generation with the other gluon
in the jet cone.

The final result for the jet vertex (see~\cite{IP12}) in the 
$(\nu,n)$-representation is free of IR and UV divergences and is
of the form $A \ln R+B+{\cal O}(R^2)$, as discussed in~\cite{Furman:1981kf}.

\section{Summary}

The NLO vertex for the forward production of a high-$p_T$ jet from an 
incoming quark or gluon, emitted by a proton, has been calculated in the SCA. 
The result has been presented in the so called $(\nu,n)$-representation, 
which turns to be very convenient for numerical implementation, as discussed
in~\cite{mesons}. 
Besides Mueller-Navelet jets, the vertex can be used also for 
forward-jet electroproduction, $\gamma^* p \to {\rm jet} + X$, in
combination with the NLO photon impact factor~\cite{chirilli}.


\begin{thebibliography}{99}

\bibitem{DGLAP}
V.N.~Gribov, L.N.~Lipatov, Sov. J. Nucl. Phys.  {\bf 15} (1972)  438;
G.~Altarelli, G.~Parisi, Nucl. Phys. B {\bf 126 } (1977) 298;
Y.L.~Dokshitzer, Sov. Phys. JETP {\bf 46 } (1977) 641.

\bibitem{BFKL}
V.S.~Fadin, E.A.~Kuraev, L.N.~Lipatov, Phys. Lett. B {\bf 60} (1975) 50;
E.A.~Kuraev, L.N.~Lipatov and V.S.~Fadin, Zh. Eksp. Teor. Fiz. {\bf 71} (1976)
840 [Sov. Phys. JETP {\bf 44} (1976) 443]; {\bf 72} (1977) 377 [{\bf 45} (1977)
199]; Ya.Ya.~Balitskii and L.N.~Lipatov, Sov. J. Nucl. Phys. {\bf 28} (1978)
822.

\bibitem{Mueller:1986ey}
A.H.~Mueller, H.~Navelet, Nucl. Phys. B {\bf 282} (1987)  727.

\bibitem{FFKP99}
V.S.~Fadin, R.~Fiore, M.I.~Kotsky and A.~Papa, Phys. Rev. D {\bf 61} (2000)
094005; Phys. Rev. D {\bf 61} (2000) 094006.

\bibitem{Ciafaloni:2000sq}
M.~Ciafaloni and G.~Rodrigo, JHEP {\bf 0005} (2000) 042.

\bibitem{Bartels:2002yj}
J.~Bartels, D.~Colferai and G.P.~Vacca, Eur. Phys. J. C {\bf 24 } (2002) 83;
Eur. Phys. J. C {\bf 29} (2003) 235;

\bibitem{Caporale:2011cc}
F.~Caporale, D.Yu.~Ivanov, B.~Murdaca, A.~Papa and A.~Perri,
JHEP {\bf 1202} (2012) 101.

\bibitem{Colferai:2010wu}
D.~Colferai, F.~Schwennsen, L.~Szymanowski, S.~Wallon,
JHEP {\bf 1012 } (2010)  026.

\bibitem{IP12}
D.Yu.~Ivanov and A.~Papa, arXiv:1202.1082.

\bibitem{Jager:2004jh}
B.~J\"ager, M.~Stratmann, W.~Vogelsang, Phys. Rev. D {\bf 70 } (2004) 034010.

\bibitem{Furman:1981kf}
M.~Furman, Nucl. Phys. B {\bf 197 } (1982) 413.

\bibitem{mesons}
D.Yu.~Ivanov and A.~Papa, Nucl. Phys. B {\bf 732} (2006) 183;
Eur. Phys. J. C {\bf 49} (2007) 947; F.~Caporale, A.~Papa and A.~Sabio Vera,
Eur. Phys. J. C {\bf 53} (2008) 525.

\bibitem{chirilli}
G.~Chirilli, these proceedings and references therein.

\end{thebibliography}
\end{document}